# AliEn Resource Brokers

Pablo Saiz
*University of the West of England, Frenchay Campus Coldharbour Lane, Bristol BS16 1QY, U.K.*
*CERN, European Organization for Nuclear Research, 1211 Geneve 23, Switzerland*

Predrag Buncic
*CERN, European Organization for Nuclear Research, 1211 Geneve 23, Switzerland*
*Institut für Kernphysik, August-Euler-Strasse 6, 60486 Frankfurt am Main, Germany*

Andreas J. Peters
*CERN, European Organization for Nuclear Research, 1211 Geneve 23, Switzerland*

for the ALICE Collaboration

AliEn (ALICE Environment) is a lightweight GRID framework developed by the Alice Collaboration. When the experiment starts running, it will collect data at a rate of approximately 2 PB per year, producing $O(10^9)$ files per year. All these files, including all simulated events generated during the preparation phase of the experiment, must be accounted and reliably tracked in the GRID environment. The backbone of AliEn is a distributed file catalogue, which associates universal logical file name to physical file names for each dataset and provides transparent access to datasets independently of physical location. The file replication and transport is carried out under the control of the File Transport Broker. In addition, the file catalogue maintains information about every job running in the system. The jobs are distributed by the Job Resource Broker that is implemented using a simplified pull (as opposed to traditional push) architecture. This paper describes the Job and File Transport Resource Brokers and shows that a similar architecture can be applied to solve both problems.

## 1. INTRODUCTION

AliEn [1] is a lightweight GRID implementation. Although originally designed for the ALICE [2] experiment, AliEn is being used by several virtual organizations. These organizations include high-energy experiments like ALICE, NA48 and NA49, and medical projects like MammoGrid [3] and GPCALMA [4].

The first component that AliEn provides is a file catalogue. The Catalogue has an interface similar to a UNIX file system, and maps logical file names (LFN) into physical file names (PFN). Files can be replicated in several locations, and the replications are done using a transfer broker.

AliEn allows the execution of jobs in the system. AliEn can be considered as a global queue system, where jobs can be executed in distributed sites in a transparent way for the end user. There is a service, called the job broker that assigns jobs to the sites where they can be executed.

This paper will explain the architecture of the AliEn resource brokers. First, an introduction to the AliEn services will be given. After that, the next two chapters will describe two of the main tasks that can be scheduled in AliEn: job execution and file replication. Next, the scheduling and execution of both tasks will be compared, given its similarities and differences. It will be followed by a more in depth discussion of the core service of the two models: the resource broker. Finally, some ideas for future work and conclusion will be exposed.

## 2. ALIEN SERVICES

AliEn components are web services, and they talk to each other using SOAP (Simple Object Access Protocol) [5]. All of the AliEn services can be seen in figure 1.

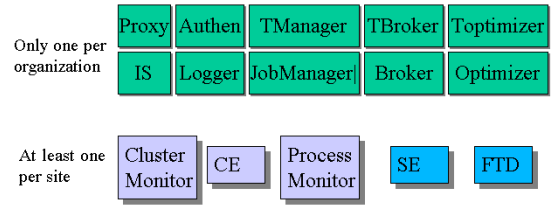

Figure 1: AliEn Services

Some of these services are unique for each virtual organisation, therefore providing a single configuration point for the management. For instance, there is only one 'Authentication' Server.

This architecture implies that there is a single point for each resource management. There are three services for each resource: a manager that keeps the status of all the tasks; a broker that assigns tasks to resources; and an optimizer, which supervises the list of waiting tasks.

## 3. ALIEN JOB EXECUTION MODEL

The AliEn job execution model is based on a central server per virtual organization, which keeps track of all the jobs that have to be executed in the system. An overview can be seen in figure 2.

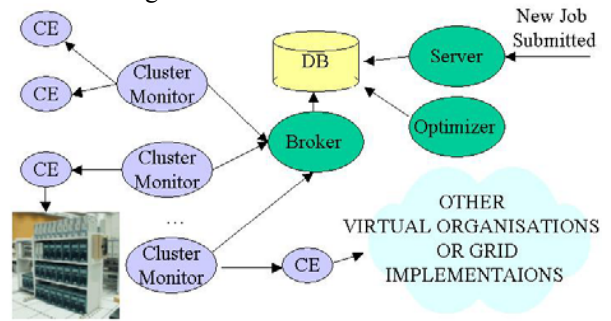

Figure 2. Computing model

## 3.1. Task: Job execution

The jobs are stored as JDL (Job Description Language) scripts. The JDL of a job specifies:
- Name of the executable that has to be run.
- Arguments to pass to the executable.
- Requirements that the worker node has to fulfill.
- Input data.
- Output data.
- Software packages.

The only field that is compulsory for the user is the name of the executable. The AliEn job manager will fill the other fields (if necessary) when a job is submitted to the system. It is also up to the job manager to convert the specifications of the user into the requirements of the job. For instance, if there is any input data, the job manager will add to the requirements a constraint specifying that the worker node has to be near the storage element that keeps the input data.

The jobs can also be prioritized. Usually, user jobs have a high priority, whereas production jobs have a lower priority and they will be executed only when there are no user jobs in the system.

## 3.2. Resource: CE

The job execution in AliEn is usually distributed over several sites. In the case of ALICE, there are more than thirty-five sites distributed over four continents. Each of these sites has at least one service called ClusterMonitor. The ClusterMonitor is used for two main reasons: first, all the connections from the site to the central services (job Manager and broker), are done through the ClusterMonitor, therefore having only one connection from each site, instead of one connection per client; furthermore, the ClusterMonitor has one or more CE (Computing Element), and it can start or stop them whenever it receives a signal.

The CE is the resource in charge of the execution of jobs. A CE is usually associated with a batch queue, and therefore can send the jobs to the worker nodes controlled by the queue. AliEn has interfaces to LSF, PBS, DQS, CONDOR and SGE. However, a CE could be associated with a single computer, in which case the jobs will be just executed in the background. A CE is defined also with a JDL, that specifies:
- Name of CE
- Hostname
- Grid Partitions to which the CE belongs.
- Storage Elements (SE) near the CE
- Software packages installed in the worker nodes.

The CE asks the Broker for jobs to execute, sending its JDL. The Broker will then try to match the JDL of the CE with the JDL of the jobs. As soon as it finds a match, it will send the job's JDL to the CE. If there are no matches, the CE will sleep for a while and ask again.

If the CE gets a job's JDL, it will send it to the batch queue, where the job will start running. The first thing that the job will do is to create a new service, called ProcessMonitor. This new web service allows the CE (and the rest of the AliEn services through the CE) to interact with the job while it is running. For example, the ProcessMonitor can send the output of the job while it is still running. The scenario of a job execution can be seen in figure 3.

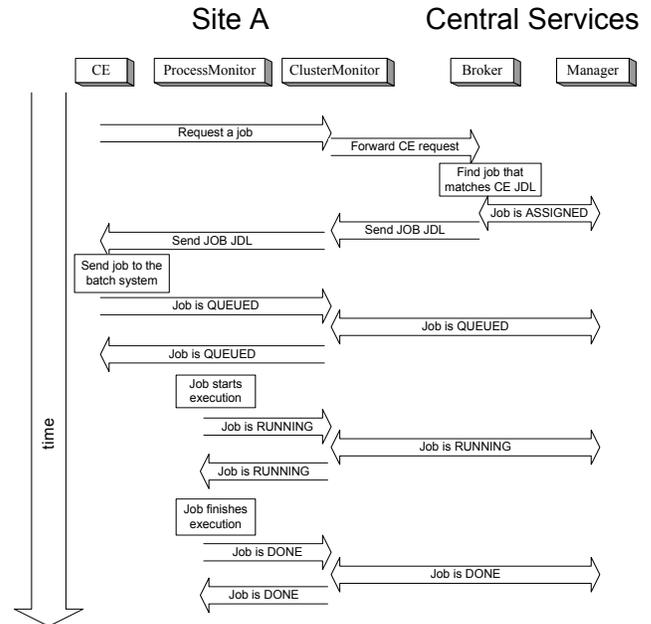

Figure 3. Job execution

A CE could be associated with another virtual organization or even with another GRID implementation. For instance, there is an interface between AliEn and EDG [6] in which AliEn sees the whole EDG as one CE.

## 3.3. Optimizer

A job optimizer checks the requirements of all the jobs waiting to be executed. The optimizer can change those requirements, therefore making it easier for the jobs to be picked up by a CE. For example, the optimizer checks if the input data has been replicated in any other SE, and modifies the constraints about the close SE accordingly. At the moment, the optimizer also suggests replication of data. In future versions, the optimizer will also suggest the installation of software packages.

## 3.4. Status flow

The different status that a job goes through when it is submitted to the system is explained in figure 4.

When a job is submitted to AliEn, its status is WAITING. A CE will pick it up and the job will be ASSIGNED, and then QUEUED when it is submitted to the local batch system. The job will start to execute on the worker node (RUNNING), and finally it will be DONE when the job finishes.

In case of failure in each of the different steps, AliEn has different job status accordingly.

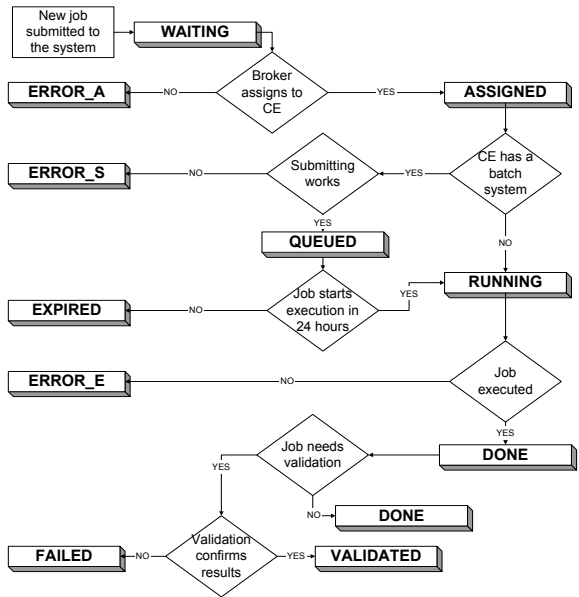

Figure 4. Job Status

Jobs can ask to be validated. The validation is a process that depends on the command that was executed. Usually, the validation parses the output of the job, checking there were no errors during the execution, and checks that the output files were produced. When a job is DONE and if the job has to be validated, a new process for the validation will start. Then, the status of the job will be VALIDATED if it passes the validation procedure, or FAILED if it does not.

## 4. ALIEN TRANSFER MODEL

The main component of the transfer model is, like in the case of the job execution model, a database with all the transfers that have to be done. The Transfer Manager is the web service responsible for inserting transfers in the system, changing their status and measuring the time each transfer spent in each state. The transfer model can be seen in figure 5.

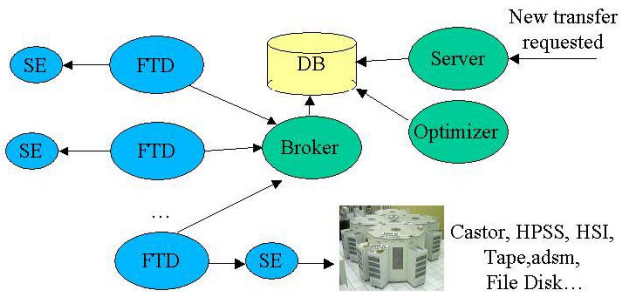

Figure 5. Transfer model

### 4.1. Task: File transfer

Transfers are specified using JDL syntax, with the following fields:
- LFN of the file to be replicated.
- SE where the file has to be replicated
- Size of the file
- Type of replication (cache, mirror or masterCopy).
- SEs that currently have the file
- Requirements of the transfer

Transfers have a priority as well. When a user requests a transfer, she has to specify the LFN, the destination SE and the type of transfer that she wants. There are three types: cache (the transferred file will not be registered in the catalogue); mirror (the new PFN will be inserted in the catalogue, and any user from the same site will get this new PFN instead of the original) and masterCopy (the PFN will be registered as the master PFN, and the previous PFN will be mirrors).

The transfer optimizer takes all the new requested transfers, and specifies the PFN and SE that currently have that LFN.

### 4.2. Resource: FTD

Each site runs at least one File Transfer Daemon (FTD). FTDs have a description in JDL format that specify:
- Name of the FTD
- SE close to the FTD.
- Free space in the cache.

Each FTD sends its JDL to the Transfer Broker, and the Broker tries to match it against the list of transfers that have to be done. If there is an action that the FTD could do, the Broker sends a JDL with the description of what the FTD is supposed to do. If there was no action, the FTD will sleep for a while and request another transfer when it wakes up. Figure 6 shows an example of a file transfer from site A to site B.

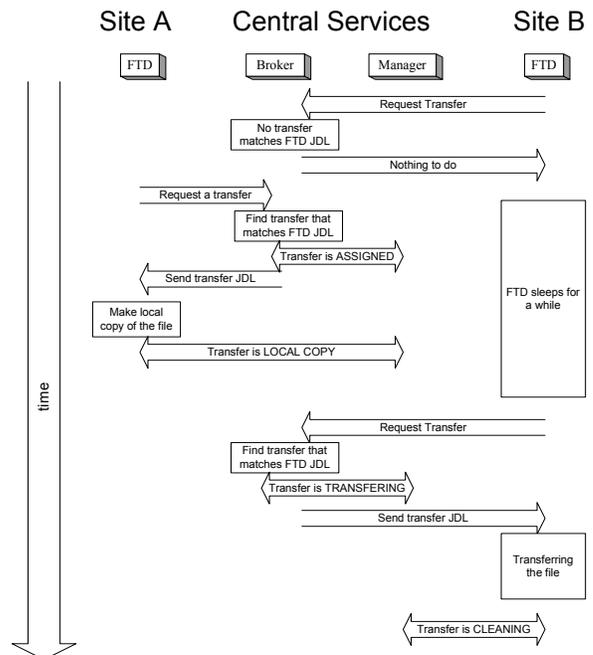

Figure 6. File transfer

At the moment, transfers are done using bbftp, a transfer software that implements its own transfer protocol which is optimized for large files [8]. However, the method to do the transfer is also specified in the JDL, therefore making it easy to try different transport protocols, or even to have different protocols depending on the target and destination of the transfer.

Like with the computing model, the transfer model makes possible to interface other grid implementation. In the AliEn-EDG interface [6], all the EDG is seen as one SE in AliEn.

### 4.3. Optimizer

The transfer optimizer has only two functions: checking for new transfers, and try to limit the number of transfer JDLs that have to be compared for an FTD.

When a transfer is scheduled, it is up to the optimizer to specify the size, current locations of the LFN and requirements.

The second function of the optimizer is to try to minimize the job of the broker. Quite often, an action needed by a transfer, can be done only by one resource. For instance, if we want to replicate a file in site A, an FTD of site A has to perform this action, and trying to match the JDL of the transfer with the JDL of an FTD of another site will be wasting the time of the broker.

### 4.4. Status flow

A transfer has to be made in several steps: first, the source FTD has to bring a copy from the local mass storage system into a scratch directory (this first step is only necessary in mass storage systems where files cannot be directly accessed by the transport mechanism); then the remote FTD can fetch the file and put it into its own storage; finally, the source FTD deletes the scratch copy of the file.

When a transfer is scheduled in the system, it goes through the status shown in figure 7.

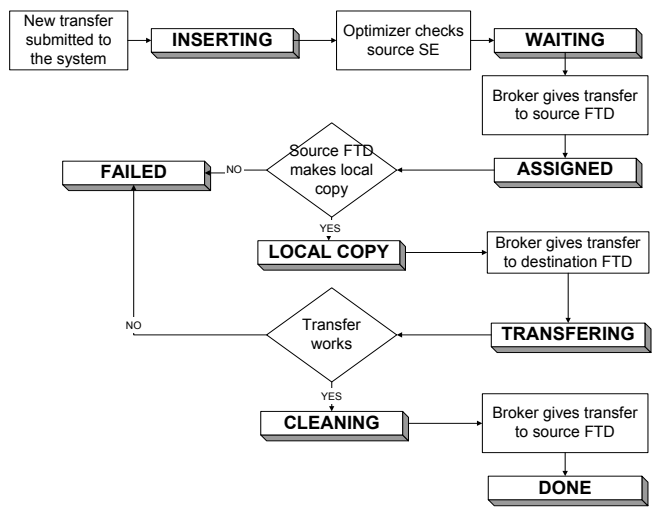

Figure 7. Transfer status

## 5. COMPARISION OF MODELS

Both the job execution and transfer models have the same structure: three central services (manager, broker and optimizer) that take care of all the tasks that have to be done, and distributed clients associated with the resources that pull the tasks that they can perform. The definition of tasks and resources is done with JDL syntax.

From the three central services, two of them do not need any information about the tasks that they are dealing with, and they could be identical for both models. The managers insert tasks in the list, change the status of the tasks, and measure how much time each task spends in each status. The brokers only care about matching the requirements of the resource with the requirements of a task. Therefore, these two services have the same design in both models.

However, the optimizers do differ for each model. Both optimizers have the same goals: to alter the requirements of the tasks so that more resources can perform them and to reduce the work necessary to find matches. But the optimizers achieve their goals in different ways.

The job optimizer could use the following strategies:
- Splitting jobs, so that each part can be submitted to different CE
- Replicating data, so that the jobs can be easily executed.
- Requesting the installation of software packages in CE.

On the other hand, the transfer optimizer has the following options:
- Grouping transfers of small files with the same source and destination to do them in one bulk.
- Redirecting the transfer routes, for instance instead of copying files directly from A to B, copy them from A to C and then from C to B.

## 6. BROKERS

The core service of both the job execution and transfer models is the broker. Both brokers have exactly the same function: receive a resource JDL, and check it against all the tasks JDL that are scheduled in the system. The brokers have the tasks ordered by priority. If there is a match between the resource and a task, the broker will send the description of the task to the resource. Otherwise, the resource sleeps for a while and tries again later. The mechanism can be seen in figure 8.

In order to do the matching, the brokers use the Condor Classads [7].

The pull architecture used in AliEn is simpler to implement than the push that it is being used in other grid systems. In a push architecture, the broker has to know the status of all the elements of the grid at all moments, so that it can decide which is the optimal resource to use for a task. Keeping the status of the whole grid is not a trivial operation. The grid is supposed to be flexible, with new elements appearing, and possible errors in the communication preventing resources from contacting the broker. In a pull architecture, if a resource goes down, the grid can go on without it. A failure in one component does not bring down the whole system.

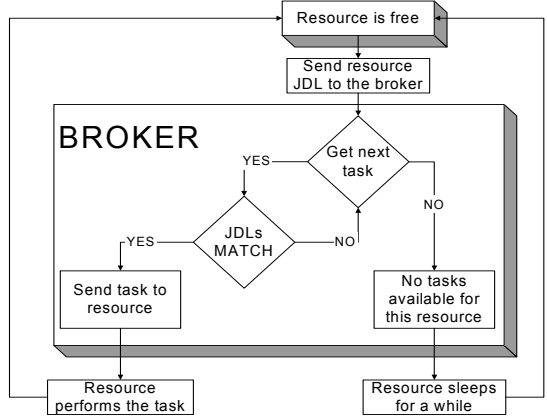

Figure 8. Resource broker

The pull architecture has proven to be stable in the more than one hundred thousand jobs that have already been executed in AliEn.

### 6.1. Broker improvements

Since the brokers are a fundamental part of the system, a great effort has been dedicated to making them robust.

The first improvement was to make the services multithreaded. During initialization, the Broker pre forks five instances of itself. Those instances will then reply to the calls from the services. The main problem with this schema was the concurrency. If there are several requests at the same time, maybe different threads will give the same task to different resources. This problem was solved locking the tables. Once a possible match has been found, the broker locks the table, checks that the task is still available, and if it is, it hands it to the resource. Finally, the broker unlocks the table.

A second improvement was done regarding the number of JDLs that the broker has to match against. As mentioned before, sometimes a task has to be performed by a resource in a specific site, and the task will never match a resource from another site. This was done adding a new column to the table of tasks. In cases when an action could only be executed by one site, this column will have the name of the site. When the broker receives a resource's JDL, it will only match it against tasks of the same site or tasks that do not specify site at all.

### 7. CONCLUSIONS

AliEn is a grid implementation that offers a distributed file catalogue and a global queue system.

The two most important tasks that can be performed in AliEn are job execution and file transfer. For each task, there is a model that fulfils it. The models are based on three central services (manager, broker and optimizer) that control the list of tasks, and on distributed services associated with resources. The distributed services ask the central broker for tasks to do, thus using a pull mechanism rather than the push.

Both the computing and transfer models in AliEn are very similar in design and implementation. The pull schema used in them proved to be an easily expandable and fault tolerant system. Several productions have already been executed, therefore proving the feasibility of the system.

The core service of both models is the broker, which assigns the tasks to the resources. The broker in a pull environment is rather simple to create, since it does not need to know the status of the grid. It does not try to find the best resource for each task, but the highest priority task that a free resource can perform.

### References

[1] P. Saiz, L. Aphecetche, P. Buncic, R. Piskac, J. -E. Revsbech and V. Sego, AliEn—ALICE environment on the GRID, **Nuclear Instruments and Methods in Physics Research Section A: Accelerators, Spectrometers, Detectors and Associated Equipment**, Volume 502, Issues 2-3 , 21 April 2003, Pages 437-440

[2] "ALICE Technical Proposal for A Large Ion Collider Experiment at the CERN LHC", CERN/LHCC/95-71, 15 December 1995.

[3] R, McClatchey on behalf of the MAMMOGRID Consortium, The MammoGrid Project Grids Architecture, these proceedings, MOAT005

[4] P. Cerello, S. Cheram, E. Lopez Torres for the GPCALMA Project and the ALICE Collaboration, Use of HEP software for medical applications, there proceedings, MOCT007

[5] SOAP Lite,  http://www.soaplite.com

[6] S.Bagnasco, R.Barbera, P.Buncic, F.Carminati, P.Cerello, P.Saiz on behalf of the ALICE Collaboration, AliEn - EDG interoperability in ALICE, these proceedings, TUCP005

[7] Condor Classified Advertisements, http://www.cs.wisc.edu/condor/classad

[8] BBFTP, http://doc.in2p3.fr/bbftp/